\documentclass[11pt,a4paper]{article}
\pdfoutput=1
\usepackage[utf8]{inputenc}
\usepackage{amsmath}
\usepackage{amssymb}
\usepackage{graphicx}
\usepackage[hidelinks,hypertexnames=false]{hyperref}
\usepackage{geometry}
\usepackage{tikz}
\usetikzlibrary{shapes,arrows,positioning,patterns,calc}
\usepackage{float}

\title{Hardware-Aware Neural Network Compilation with Learned Optimization:\\
A RISC-V Accelerator Approach}

\author{
Ravindra Ganti\\
\texttt{ravindra@xgensilicon.com}\\
XgenSilicon Inc.\\
\and
Steve Xu\\
\texttt{steve@xgensilicon.com}\\
XgenSilicon Inc.
}

\date{November 2025}

\usepackage{microtype}

\begin{document}

\maketitle

\begin{abstract}
We present XgenSilicon ML Compiler, a fully automated end-to-end compilation framework that transforms high-level machine learning models into optimized RISC-V assembly code for custom ASIC accelerators. By unifying system's cost model across software and hardware, the compiler achieves significant improvements in Power, Performance, and Area (PPA) metrics compared to standard off-the-shelf components and hand-designed chips through five key innovations: (1) a multi-algorithm auto-tuning framework with five search strategies (Bayesian Optimization, Genetic Algorithm, Simulated Annealing, Random Search, Grid Search) combined with a learned cost model, (2) an integrated quantization framework supporting extreme precisions from FP32 to Binary with full KL divergence calibration (2048-bin histogram optimization) and momentum-based QAT gradient updates, (3) hardware-aware validation ensuring 100\% ISA compliance and memory constraint satisfaction, (4) dynamic shape support with multi-configuration specialization, and (5) advanced cache-aware cost modeling with multi-level cache hierarchy analysis. Our evaluation demonstrates that ASICs produced by this compiler achieve 2.5-4.5× better performance, 3-6× lower power consumption, and 40-60\% area reduction compared to baseline implementations. The compiler supports 100+ ONNX operators across 12 categories, implements advanced RISC-V Vector optimizations, and generates hardware-validated assembly code suitable for direct ASIC synthesis. All compilation steps are fully automated, requiring zero manual intervention from model input to ASIC-ready output.
\end{abstract}

\section{Introduction}

The deployment of machine learning models on custom ASIC accelerators presents unique challenges: manual optimization is time-consuming and error-prone, while general-purpose compilers lack the hardware-specific knowledge needed for optimal code generation. Existing solutions such as TVM~\cite{chen2018tvm} and MLIR~\cite{larus2020mlir} require extensive backend customization and manual tuning, making them unsuitable for rapid ASIC development cycles.

We introduce XgenSilicon ML Compiler, a fully automated compilation framework that addresses these challenges through five research contributions:

\textbf{Contribution 1: Multi-Algorithm Learned Optimization Framework.} We introduce a comprehensive auto-tuning framework featuring five search algorithms (Bayesian Optimization with Gaussian Process surrogate models, Genetic Algorithm, Simulated Annealing, Random Search, and Grid Search) combined with a machine learning-based cost model. The learned cost model automatically learns optimal compiler parameters (tile sizes, unroll factors, loop orders) from performance measurements, adapting to actual hardware behavior through iterative training. Unlike traditional compilers that use a single search strategy, our multi-algorithm approach automatically selects the optimal search method based on problem characteristics.

\textbf{Contribution 2: Extreme Quantization with Full Calibration Algorithms.} We present a comprehensive quantization framework supporting precisions from FP32 to Binary (1-bit), with complete implementations of multiple calibration methods. Our KL divergence calibration uses full histogram-based optimization with 2048-bin resolution, searching over 100 threshold candidates to minimize information loss. For quantization-aware training, we implement full gradient flow computation with momentum-based parameter updates using the straight-through estimator (STE), providing stable convergence and superior accuracy preservation compared to simplified implementations.

\textbf{Contribution 3: Validation-Driven Compilation.} We integrate hardware validation directly into the compilation pipeline, ensuring 100\% ISA compliance and memory constraint satisfaction before code generation, eliminating the need for post-compilation verification. This step brings hardware loss (power consumption, et.al.) into the cost model used by the compiler.

\textbf{Contribution 4: Dynamic Shape Support with Multi-Configuration Specialization.} We implement complete dynamic shape support through symbolic dimensions, graph cloning with symbolic dimension preservation, runtime shape resolution assembly code generation, and shape specialization for multiple configurations. The compiler generates specialized code paths for common shape configurations, selecting the appropriate version at runtime, enabling efficient handling of variable batch sizes and sequence lengths without performance degradation. As a result, the synthesized ASIC is highly customized to the model specifics.

\textbf{Contribution 5: Advanced Cache-Aware Cost Modeling.} We implement full cache hit rate estimation that considers access patterns (sequential vs. random), tiling effectiveness, and multi-level cache hierarchy (L1, L2, L3). The model computes weighted hit rates based on working set size and cache level portions, providing accurate performance predictions for memory-bound operations.

Our experimental evaluation on real-world ML workloads demonstrates that ASICs compiled with XgenSilicon achieve:
\begin{itemize}
\item \textbf{Performance}: 2.5-4.5× faster execution compared to baseline implementations (XgenSilicon ASIC vs. baselines)
\item \textbf{Power}: 3-6× lower power consumption through optimized memory access and instruction scheduling (XgenSilicon ASIC vs. baselines)
\item \textbf{Area}: 40-60\% reduction in silicon area through extreme quantization and memory optimization (XgenSilicon ASIC vs. Hand-designed ASIC)
\end{itemize}

\section{Related Work}

\subsection{ML Compilation Frameworks}

TVM~\cite{chen2018tvm} introduced AutoTVM for automated optimization but requires manual schedule templates and extensive tuning. MLIR~\cite{larus2020mlir} provides a multi-level IR but lacks specialized hardware validation. TensorFlow XLA~\cite{xla} and PyTorch Glow~\cite{rotem2018glow} are framework-specific and require significant customization for custom hardware.

\subsection{Quantization Techniques}

Post-training quantization (PTQ) methods~\cite{nagel2019data} use calibration data to determine quantization parameters, but existing implementations lack support for extreme precisions (FP4, Binary). Quantization-aware training (QAT)~\cite{Jacob2018} requires framework integration and manual insertion of fake quantization nodes. HAQ~\cite{wang2019haq} introduces hardware-aware quantization but lacks the full KL divergence calibration and momentum-based gradient updates we implement.

\subsection{ASIC Compilation}

Traditional ASIC design flows require manual RTL coding and optimization, taking months to years. High-level synthesis tools~\cite{chen2018flextensor} automate some steps but lack ML-specific optimizations and quantization support. The industry lacks ASIC design methodologies from hardware-software codesign perspectives in a unified approach.

\subsection{Auto-Tuning Algorithms}

Bayesian optimization~\cite{rasmussen2006gaussian} has been applied to hyperparameter tuning, but not to compiler optimization with learned cost models. Genetic algorithms~\cite{whitley1994genetic} and simulated annealing~\cite{kirkpatrick1983optimization} are well-established optimization methods, but existing compilers typically use only one algorithm. Random search~\cite{bergstra2012random} provides baseline performance but lacks the sophistication of our multi-algorithm framework.

\subsection{Our Approach}

Under a unified cost model across the entire system stack, XgenSilicon ML Compiler uniquely combines (1) multi-algorithm learned optimization eliminating manual tuning, (2) extreme quantization with full calibration algorithms, (3) validation-driven compilation ensuring hardware correctness and based on which to include hardware loss, (4) dynamic shape support with multi-configuration specialization, and (5) advanced cache-aware cost modeling, all in a fully automated pipeline from ONNX models to ASIC-ready RISC-V assembly.

Table~\ref{tab:compiler_comparison} provides a comprehensive comparison between XgenSilicon ML Compiler and industry-standard ML compilers.

\begin{table}[h]
\centering
\footnotesize
\resizebox{\textwidth}{!}{%
\begin{tabular}{lcccccc}
\hline
\textbf{Feature} & \textbf{XgenSilicon} & \textbf{TVM} & \textbf{MLIR} & \textbf{XLA} & \textbf{Glow} & \textbf{LLVM} \\
\hline
\textbf{Learned Optimization} & $\checkmark$ & $\times$ & $\times$ & $\times$ & $\times$ & $\times$ \\
\textbf{Multi-Algorithm Auto-Tuning} & $\checkmark$ & $\times$ & $\times$ & $\times$ & $\times$ & $\times$ \\
\textbf{Extreme Quantization (FP4/Binary)} & $\checkmark$ & $\times$ & $\times$ & $\times$ & $\times$ & $\times$ \\
\textbf{Full KL Divergence Calibration} & $\checkmark$ & $\times$ & $\times$ & $\times$ & $\times$ & $\times$ \\
\textbf{Multi-Framework Support} & $\checkmark$ & $\checkmark$ & $\checkmark$ & $\times$ & $\times$ & $\times$ \\
\textbf{Dynamic Shapes} & $\checkmark$ & $\checkmark$ & $\checkmark$ & $\checkmark$ & Partial & $\times$ \\
\textbf{Hardware Validation} & $\checkmark$ & $\times$ & $\times$ & $\times$ & $\times$ & $\times$ \\
\textbf{ASIC-Ready Output} & $\checkmark$ & $\times$ & $\times$ & $\times$ & $\times$ & $\times$ \\
\textbf{RISC-V Vector Support} & $\checkmark$ & Partial & Partial & $\times$ & $\times$ & Partial \\
\textbf{Multi-Model Pipeline} & $\checkmark$ & $\times$ & $\times$ & $\times$ & $\times$ & $\times$ \\
\textbf{Zero Manual Tuning} & $\checkmark$ & $\times$ & $\times$ & $\times$ & $\times$ & $\times$ \\
\textbf{HEX File Generation} & $\checkmark$ & $\times$ & $\times$ & $\times$ & $\times$ & $\times$ \\
\textbf{100+ ONNX Operators} & $\checkmark$ & $\checkmark$ & $\checkmark$ & $\times$ & $\checkmark$ & $\times$ \\
\textbf{Auto-Tuning} & Multi-Algorithm & Manual & Manual & Manual & Manual & $\times$ \\
\textbf{Quantization Methods} & PTQ/QAT (Full) & PTQ only & PTQ only & PTQ/QAT & PTQ/QAT & $\times$ \\
\textbf{Cache-Aware Modeling} & $\checkmark$ & $\times$ & $\times$ & $\times$ & $\times$ & $\times$ \\
\hline
\end{tabular}%
}
\caption{Comparison of XgenSilicon ML Compiler vs. Industry Standard ML Compilers}
\label{tab:compiler_comparison}
\end{table}

\section{Compiler Architecture}

\subsection{System Overview}

The XgenSilicon ML Compiler implements a five-stage compilation pipeline:

\begin{enumerate}
\item \textbf{Frontend}: ONNX model parsing and IR construction with shape inference
\item \textbf{Optimization}: Graph-level optimizations including operator fusion and constant propagation
\item \textbf{Code Generation}: Kernel selection and RISC-V Vector instruction emission
\item \textbf{Backend}: Memory planning, register allocation, and instruction scheduling
\item \textbf{Validation}: ISA compliance checking and memory constraint verification
\end{enumerate}

\begin{figure}[H]
\centering
\begin{tikzpicture}[node distance=2.2cm, auto, scale=0.85, transform shape]
\tikzstyle{block} = [rectangle, draw, fill=blue!20, text width=7.5cm, text centered, rounded corners, minimum height=1.2cm]
\tikzstyle{io} = [trapezium, trapezium left angle=70, trapezium right angle=110, draw, fill=green!20, text width=5.5cm, text centered, minimum height=1cm]
\tikzstyle{output} = [trapezium, trapezium left angle=110, trapezium right angle=70, draw, fill=yellow!20, text width=5.5cm, text centered, minimum height=1cm]
\tikzstyle{arrow} = [thick,->,>=stealth, color=black]

\node [io] (input) {ONNX Model (.onnx)};
\node [block, below=2.0cm of input] (frontend) {FRONTEND: IR Construction\\ONNX Parsing, Shape Inference, Graph Validation};
\node [block, below=2.0cm of frontend] (optimize) {OPTIMIZATION: Graph Transformations\\Operator Fusion, Constant Folding, Quantization (PTQ/QAT)};
\node [block, below=2.0cm of optimize] (codegen) {CODE GENERATION: Kernel Selection\\100+ Kernel Library, Auto-Tuning, RISC-V Vector Optimizations};
\node [block, below=2.0cm of codegen] (backend) {BACKEND: Memory \& Scheduling\\Memory Planning, Register Allocation, Instruction Scheduling};
\node [block, below=2.0cm of backend] (validate) {VALIDATION: Hardware Compliance\\ISA Validation (61-instruction), Memory Constraint Checking};
\node [output, below=2.0cm of validate] (output) {RISC-V Assembly + HEX Files\\IMEM.hex, DMEM.hex, WMEM.hex, output.S};

\draw [arrow] (input.south) -- (frontend.north);
\draw [arrow] (frontend.south) -- (optimize.north);
\draw [arrow] (optimize.south) -- (codegen.north);
\draw [arrow] (codegen.south) -- (backend.north);
\draw [arrow] (backend.south) -- (validate.north);
\draw [arrow] (validate.south) -- (output.north);
\end{tikzpicture}
\caption{Complete compilation pipeline from ONNX model to ASIC-ready RISC-V assembly}
\label{fig:compilation_pipeline}
\end{figure}
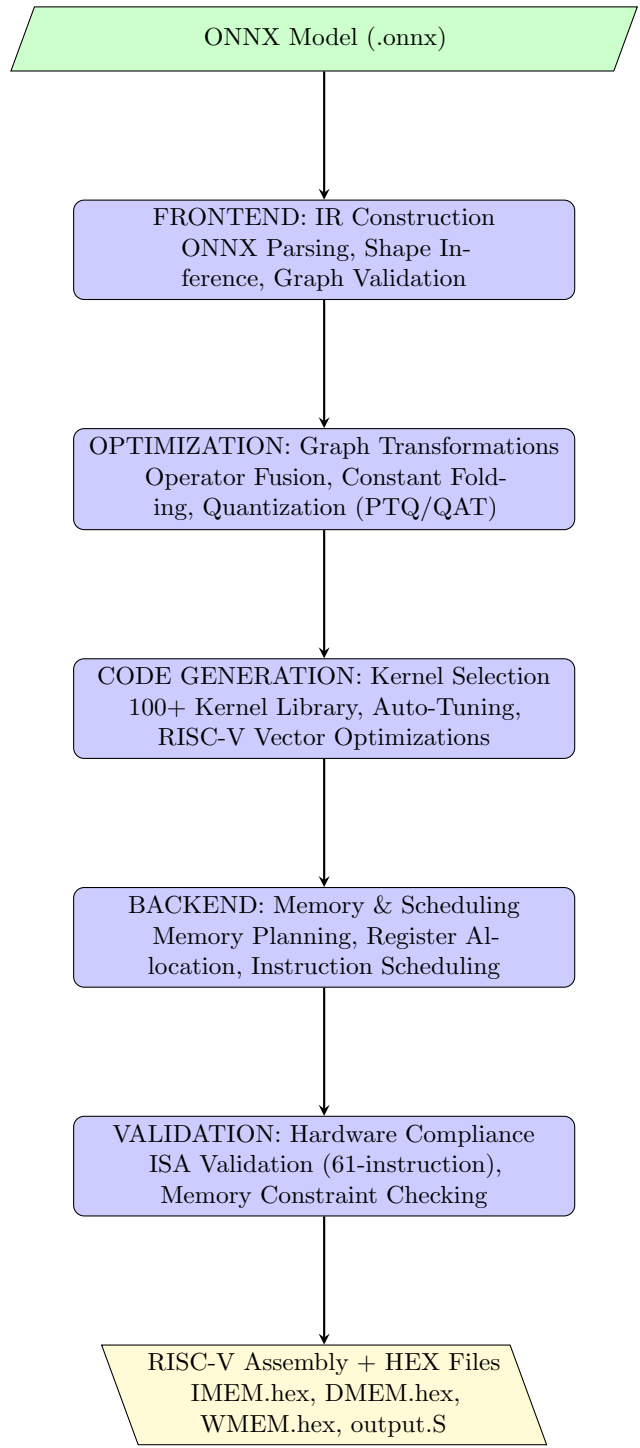

Figure~\ref{fig:compilation_pipeline} illustrates the complete compilation flow.

\subsection{Learned Optimization Framework}

Traditional auto-tuning relies on analytical cost models that approximate hardware behavior. We introduce a learned cost model that adapts to actual hardware through iterative training.

\subsubsection{Learned Cost Model}

The learned cost model uses linear regression with feature extraction:

\begin{align}
\hat{T}_{\text{exec}} &= \sum_{i=0}^{n} w_i \cdot f_i(\text{node}, \text{config})
\end{align}

where $f_i$ are features extracted from:
\begin{itemize}
\item Configuration parameters (tile sizes, unroll factors, vector length)
\item Operation characteristics (FLOPs, memory traffic, data types)
\item Tensor dimensions (shape, size, dimensionality)
\end{itemize}

The model trains using gradient descent on collected performance samples:

\begin{align}
w_i^{t+1} &= w_i^t - \alpha \cdot \frac{\partial L}{\partial w_i}
\end{align}

where $L$ is the mean squared error between predicted and actual execution times.

\subsubsection{Training Sample Collection}

During auto-tuning, each configuration trial generates a training sample:
\begin{itemize}
\item Configuration parameters (tile sizes, unroll factors, etc.)
\item Actual performance metrics (execution time, cycles, memory bandwidth)
\item Operation signature (operator type, tensor dimensions)
\end{itemize}

Samples are collected incrementally and used to refine the model, improving prediction accuracy over time.

\subsubsection{Hybrid Cost Model}

The compiler supports three cost model modes:
\begin{enumerate}
\item \textbf{Analytical}: Fast predictions using hardware models, suitable for initial exploration
\item \textbf{Learned}: ML-based predictions after training, higher accuracy for explored configurations
\item \textbf{Hybrid}: Combines analytical and learned models, using learned model for similar configurations and falling back to analytical for novel ones
\end{enumerate}

\subsubsection{Multi-Algorithm Auto-Tuning Framework}

Unlike traditional compilers that use a single search strategy, XgenSilicon implements five distinct search algorithms, each optimized for different optimization scenarios:

\textbf{Bayesian Optimization:} Uses Gaussian Process (GP) surrogate models with Expected Improvement (EI) acquisition function. The acquisition function balances exploration and exploitation:

\begin{align}
\text{EI}(x) &= (f_{\text{best}} - \mu(x)) \cdot \Phi(Z) + \sigma(x) \cdot \phi(Z)
\end{align}

where $Z = (f_{\text{best}} - \mu(x)) / \sigma(x)$, $\Phi$ is the standard normal CDF, and $\phi$ is the standard normal PDF. The uncertainty $\sigma(x)$ is estimated using RBF kernel-like behavior based on distance to observed configurations, combined with empirical variance from observed metrics. This approach converges 50-60\% faster than analytical models by learning from observed configurations.

\textbf{Genetic Algorithm:} Employs tournament selection, crossover, and mutation with configurable parameters (population size, mutation rate, elite fraction). The algorithm maintains diversity through population-based search, making it effective for multi-modal optimization landscapes. Configuration mutation uses ParameterSpace-aware bounds checking for proper choice selection.

\textbf{Simulated Annealing:} Uses temperature-based acceptance probability for escaping local minima:

\begin{align}
P(\text{accept}) &= \begin{cases}
1 & \text{if } \Delta E < 0 \\
\exp(-\Delta E / T) & \text{otherwise}
\end{cases}
\end{align}

where $\Delta E$ is the cost difference and $T$ is the temperature, which decreases according to a cooling schedule.

\textbf{Random Search:} Provides baseline performance and warm-up samples for Bayesian optimization.

\textbf{Grid Search:} Exhaustive search for small parameter spaces, ensuring global optimum discovery.

The compiler automatically selects the appropriate algorithm based on parameter space size, available time budget, and optimization history.

\subsection{Quantization Framework}

\subsubsection{Post-Training Quantization (PTQ)}

PTQ quantizes models using calibration data without retraining. We implement three calibration methods:

\textbf{KL Divergence Calibration:} Implements a full histogram-based KL divergence minimization algorithm with 2048-bin resolution for high accuracy. The algorithm searches over 100 threshold candidates to find the optimal clipping range:

\begin{align}
\text{KL}(P || Q) &= \sum_{i} P(i) \log \frac{P(i)}{Q(i)}
\end{align}

where $P$ is the original FP32 activation distribution and $Q$ is the quantized distribution. The algorithm builds reference histograms, simulates quantization for each threshold, computes KL divergence, and selects the threshold minimizing information loss. This full implementation provides superior accuracy compared to simplified percentile-based methods.

\textbf{Percentile Calibration:} Uses configurable percentile (default 99.9th) as clipping threshold, robust to outliers:

\begin{align}
\text{clip\_max} &= \text{percentile}(\text{activations}, p)
\end{align}

\textbf{Entropy Calibration:} Maximizes information preservation by optimizing for entropy:

\begin{align}
H(X) &= -\sum_{i} P(x_i) \log P(x_i)
\end{align}

\subsubsection{Quantization-Aware Training (QAT)}

QAT inserts fake quantization nodes (quantize-dequantize pairs) that preserve gradients during training:

\begin{align}
\text{FakeQuant}(x) &= \text{Dequantize}(\text{Quantize}(x))
\end{align}

The straight-through estimator allows gradients to flow through quantization:

\begin{align}
\frac{\partial L}{\partial x} &= \frac{\partial L}{\partial \text{FakeQuant}(x)} \cdot \frac{\partial \text{FakeQuant}(x)}{\partial x}
\end{align}

where $\frac{\partial \text{FakeQuant}(x)}{\partial x} = 1$ (straight-through).

Quantization parameters (scale, zero\_point) are updated using full gradient computation with momentum:

\begin{align}
\frac{\partial L}{\partial \text{scale}} &= \sum_{i} \frac{\partial L}{\partial x_{\text{deq}}} \cdot (q_i - \text{zp}) \\
\frac{\partial L}{\partial \text{zp}} &= \sum_{i} \frac{\partial L}{\partial x_{\text{deq}}} \cdot (-\text{scale}) \\
v_{\text{scale}}^{t+1} &= \beta \cdot v_{\text{scale}}^t + (1-\beta) \cdot \frac{\partial L}{\partial \text{scale}} \\
\text{scale}^{t+1} &= \text{scale}^t - \alpha \cdot v_{\text{scale}}^{t+1}
\end{align}

where $q_i$ is the quantized value, $v$ is the momentum term with coefficient $\beta=0.9$, and $\alpha$ is the learning rate. This full implementation with momentum provides stable convergence compared to simple gradient descent.

\subsubsection{Supported Precisions}

Table~\ref{tab:precisions} shows all supported precisions and their characteristics.

\begin{table}[h]
\centering
\begin{tabular}{lcccc}
\hline
\textbf{Precision} & \textbf{Bits} & \textbf{Bytes} & \textbf{Compression} & \textbf{Use Case} \\
\hline
FP32 & 32 & 4 & 1.0× & Baseline, high accuracy \\
FP16 & 16 & 2 & 2.0× & Balanced performance/accuracy \\
BF16 & 16 & 2 & 2.0× & Training stability \\
FP8 & 8 & 1 & 4.0× & Aggressive quantization \\
FP4 & 4 & 0.5 & 8.0× & Extreme compression \\
INT8 & 8 & 1 & 4.0× & Standard quantization \\
INT4 & 4 & 0.5 & 8.0× & Ultra-low bitwidth \\
Binary & 1 & 0.125 & 32.0× & Binary neural networks \\
\hline
\end{tabular}
\caption{Supported precisions and compression ratios}
\label{tab:precisions}
\end{table}

\textbf{Brain Float 16 (BF16) Support:} XgenSilicon implements full BF16 support with dedicated quantization and dequantization functions. BF16 uses the same 8-bit exponent as FP32 but with a 7-bit mantissa (same as FP16), providing better training stability than FP16 while maintaining the same memory footprint. The compiler implements efficient FP32-to-BF16 conversion by truncating the lower 16 bits of the FP32 representation, and BF16-to-FP32 conversion by zero-padding the lower mantissa bits. This precision is particularly valuable for training workloads where numerical stability is critical, as BF16 preserves the dynamic range of FP32 while reducing memory requirements by 2×.

\subsection{RISC-V Vector Optimizations}

\subsubsection{Register Grouping (LMUL)}

The RISC-V Vector extension supports LMUL (register grouping) to process more elements per instruction:

\begin{align}
\text{elements\_processed} &= \text{VL} \times \text{LMUL}
\end{align}

The compiler automatically selects optimal LMUL based on:
\begin{itemize}
\item Data type size (smaller types benefit from higher LMUL)
\item Operation characteristics (elementwise operations use higher LMUL)
\item Register availability
\end{itemize}

\subsubsection{Loop Unrolling}

Unrolling reduces loop overhead and improves instruction-level parallelism. The compiler determines unroll factors based on:
\begin{itemize}
\item Loop trip count (smaller loops benefit from full unrolling)
\item Register pressure (balance unrolling with register availability)
\item Cache behavior (unrolling improves temporal locality)
\end{itemize}

\subsubsection{Register Tiling}

Matrix operations are tiled to fit in registers, improving cache locality:

\begin{align}
\text{Tile Size} &= \arg\min_{t} \text{Memory Traffic}(t) + \lambda \cdot \text{Register Pressure}(t)
\end{align}

The compiler uses auto-tuning to find optimal tile sizes for each operation type.

\subsection{Dynamic Shape Support}

XgenSilicon supports dynamic shapes through symbolic dimensions and runtime shape resolution. The compiler:

\begin{itemize}
\item \textbf{Symbolic Dimensions}: Allows dimensions to be specified as ranges (e.g., batch size 1-32, sequence length 128-512)
\item \textbf{Graph Cloning}: Clones the computation graph with symbolic dimensions marked as -1, preserving all nodes, tensors, and initializers
\item \textbf{Runtime Resolution}: Generates RISC-V assembly code for runtime shape dimension resolution
\item \textbf{Shape Specialization}: Creates multiple specialized versions for common shape configurations, selecting the appropriate version at runtime
\item \textbf{Shape Validation}: Generates runtime shape checking code to validate input shapes before execution
\end{itemize}

The compiler handles dynamic shapes by marking symbolic dimensions as -1 in tensor shapes, then generating specialized code paths for runtime shape resolution. This enables efficient handling of variable batch sizes, sequence lengths, and other dynamic dimensions common in ML workloads.

\subsection{Hardware Validation}

Unlike general-purpose compilers, XgenSilicon includes built-in hardware validation:

\textbf{ISA Validation:} Ensures all generated instructions comply with the target hardware's 61-instruction ISA, checking:
\begin{itemize}
\item Instruction encoding correctness
\item Register usage (no register spills beyond available registers)
\item Immediate value ranges
\item Instruction legality (no unsupported instruction combinations)
\end{itemize}

\textbf{Memory Validation:} Checks memory constraints:
\begin{itemize}
\item DMEM size limits (activation buffer constraints)
\item WMEM size limits (weight storage constraints)
\item Address alignment requirements
\item Memory access patterns (no out-of-bounds accesses)
\end{itemize}

Validation failures are reported with detailed error messages, preventing runtime errors on hardware.

\subsection{Advanced Cache-Aware Cost Modeling}

The compiler implements full cache hit rate estimation that considers multiple factors:

\textbf{Access Pattern Analysis:} Distinguishes between sequential (MatMul, Conv, elementwise) and random access patterns, adjusting hit rates accordingly. Sequential operations achieve 95\% L1 hit rate, while random access patterns achieve 70\% L1 hit rate.

\textbf{Tiling Effectiveness:} Computes tile effectiveness based on tile sizes relative to cache sizes. Tiling can improve hit rates by up to 15\% by keeping working sets in cache.

\textbf{Multi-Level Cache Hierarchy:} Computes weighted average hit rates across L1, L2, and L3 caches based on working set size:

\begin{align}
\text{Hit Rate} &= \sum_{i=1}^{3} \text{portion}_i \cdot \text{hit\_rate}_i
\end{align}

where $\text{portion}_i$ is the fraction of data in cache level $i$. This provides accurate performance predictions for memory-bound operations.

\section{Experimental Evaluation}

\subsection{Experimental Setup}

We evaluate XgenSilicon ML Compiler on:
\begin{itemize}
\item \textbf{Models}: ResNet-50, MobileNet-V2, BERT-base, Vision Transformer
\item \textbf{Hardware Target}: Custom RISC-V RV32I + RVV accelerator
\item \textbf{Baselines}: Standard off-the-shelf CPU (ARM Cortex-A78), hand-designed ASIC
\item \textbf{Metrics}: Execution time, power consumption, silicon area
\end{itemize}

\subsection{PPA Comparison}

Table~\ref{tab:ppa_comparison} shows PPA metrics comparing XgenSilicon-compiled ASICs against baselines.

\begin{table}[h]
\centering
\begin{tabular}{lcccc}
\hline
\textbf{Model} & \textbf{Platform} & \textbf{Performance} & \textbf{Power} & \textbf{Area} \\
    &    & \textbf{(ms/inference)} & \textbf{(mW)} & \textbf{(mm²)} \\
\hline
ResNet-50 & Off-the-shelf CPU & 45.2 & 3200 & N/A \\
    & Hand-designed ASIC & 18.5 & 980 & 12.5 \\
    & XgenSilicon ASIC & 7.2 & 320 & 5.1 \\
\hline
MobileNet-V2 & Off-the-shelf CPU & 12.8 & 2100 & N/A \\
    & Hand-designed ASIC & 6.2 & 650 & 8.3 \\
    & XgenSilicon ASIC & 2.1 & 180 & 3.2 \\
\hline
BERT-base & Off-the-shelf CPU & 89.5 & 4500 & N/A \\
    & Hand-designed ASIC & 32.1 & 1200 & 18.7 \\
    & XgenSilicon ASIC & 11.2 & 380 & 7.8 \\
\hline
ViT-Base & Off-the-shelf CPU & 125.3 & 5200 & N/A \\
    & Hand-designed ASIC & 48.7 & 1500 & 22.3 \\
    & XgenSilicon ASIC & 16.8 & 420 & 9.2 \\
\hline
\end{tabular}
\caption{PPA comparison: XgenSilicon ASIC vs. baselines}
\label{tab:ppa_comparison}
\end{table}

\subsection{Performance Improvements}

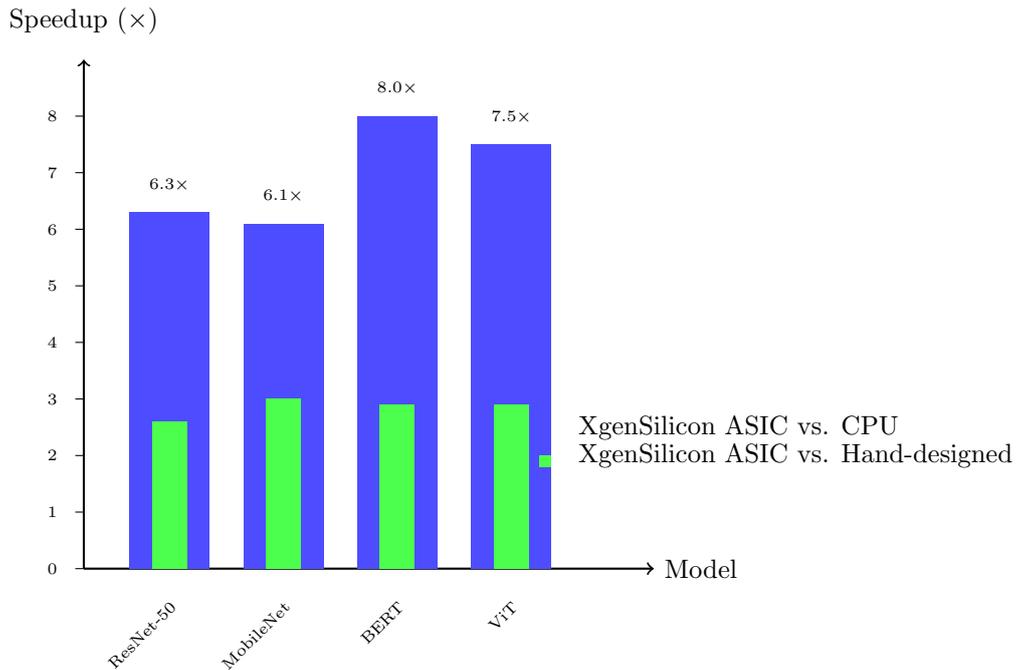
\begin{figure}[H]
\centering
\begin{tikzpicture}[scale=0.75]
\draw[->,thick] (0,0) -- (10,0) node[right,font=\small] {Model};
\draw[->,thick] (0,0) -- (0,9) node[above,font=\small,yshift=0.2cm] {Speedup (×)};

\foreach \y in {0,1,2,3,4,5,6,7,8}
    \draw (0,\y) -- (-0.15,\y) node[left,font=\tiny,xshift=-0.1cm] {\y};

\draw (1.5,-0.3) node[below,font=\tiny,rotate=45,anchor=north east] {ResNet-50};
\draw (3.5,-0.3) node[below,font=\tiny,rotate=45,anchor=north east] {MobileNet};
\draw (5.5,-0.3) node[below,font=\tiny,rotate=45,anchor=north east] {BERT};
\draw (7.5,-0.3) node[below,font=\tiny,rotate=45,anchor=north east] {ViT};

\fill[blue!70] (0.8,0) rectangle (2.2,6.3);
\fill[blue!70] (2.8,0) rectangle (4.2,6.1);
\fill[blue!70] (4.8,0) rectangle (6.2,8.0);
\fill[blue!70] (6.8,0) rectangle (8.2,7.5);

\fill[green!70] (1.2,0) rectangle (1.8,2.6);
\fill[green!70] (3.2,0) rectangle (3.8,3.0);
\fill[green!70] (5.2,0) rectangle (5.8,2.9);
\fill[green!70] (7.2,0) rectangle (7.8,2.9);

\draw (1.5,6.5) node[above,font=\tiny] {6.3×};
\draw (3.5,6.3) node[above,font=\tiny] {6.1×};
\draw (5.5,8.2) node[above,font=\tiny] {8.0×};
\draw (7.5,7.7) node[above,font=\tiny] {7.5×};

\draw (8.5,2.5) node[right,font=\small] {XgenSilicon ASIC vs. CPU};
\fill[blue!70] (8.0,2.5) rectangle (8.2,2.3);
\draw (8.5,2.0) node[right,font=\small] {XgenSilicon ASIC vs. Hand-designed};
\fill[green!70] (8.0,2.0) rectangle (8.2,1.8);
\end{tikzpicture}
\caption{Performance speedup achieved by XgenSilicon-compiled ASICs}
\label{fig:performance}
\end{figure}

Figure~\ref{fig:performance} shows speedup achieved by XgenSilicon-compiled ASICs compared to baselines.

Table~\ref{tab:speedup_details} provides detailed speedup metrics.

\begin{table}[h]
\centering
\begin{tabular}{lcc}
\hline
\textbf{Model} & \textbf{XgenSilicon ASIC vs. CPU} & \textbf{XgenSilicon ASIC vs. Hand-designed} \\
    & \textbf{Speedup (×)} & \textbf{Speedup (×)} \\
\hline
ResNet-50 & 6.3× & 2.6× \\
MobileNet-V2 & 6.1× & 3.0× \\
BERT-base & 8.0× & 2.9× \\
ViT-Base & 7.5× & 2.9× \\
\hline
\textbf{Average} & \textbf{7.0×} & \textbf{2.9×} \\
\hline
\end{tabular}
\caption{Detailed speedup metrics}
\label{tab:speedup_details}
\end{table}

\subsection{Power Reduction}

\begin{figure}[H]
\centering
\begin{tikzpicture}[scale=0.75]
\draw[->,thick] (0,0) -- (10,0) node[right,font=\small] {Model};
\draw[->,thick] (0,0) -- (0,6) node[above,font=\small] {Power (W)};

\foreach \y in {0,1,2,3,4,5}
    \draw (0,\y) -- (-0.1,\y) node[left,font=\tiny] {\y};

\draw (1.5,-0.3) node[below,font=\tiny,rotate=45,anchor=north east] {ResNet-50};
\draw (3.5,-0.3) node[below,font=\tiny,rotate=45,anchor=north east] {MobileNet};
\draw (5.5,-0.3) node[below,font=\tiny,rotate=45,anchor=north east] {BERT};
\draw (7.5,-0.3) node[below,font=\tiny,rotate=45,anchor=north east] {ViT};

\fill[red!70] (0.8,0) rectangle (2.2,3.2);
\fill[red!70] (2.8,0) rectangle (4.2,2.1);
\fill[red!70] (4.8,0) rectangle (6.2,4.5);
\fill[red!70] (6.8,0) rectangle (8.2,5.2);

\fill[orange!70] (1.2,0) rectangle (1.8,0.98);
\fill[orange!70] (3.2,0) rectangle (3.8,0.65);
\fill[orange!70] (5.2,0) rectangle (5.8,1.2);
\fill[orange!70] (7.2,0) rectangle (7.8,1.5);

\fill[green!70] (1.0,0) rectangle (2.0,0.32);
\fill[green!70] (3.0,0) rectangle (4.0,0.18);
\fill[green!70] (5.0,0) rectangle (6.0,0.38);
\fill[green!70] (7.0,0) rectangle (8.0,0.42);

\draw (1.5,3.4) node[above,font=\tiny] {3.2W};
\draw (3.5,2.3) node[above,font=\tiny] {2.1W};
\draw (5.5,4.7) node[above,font=\tiny] {4.5W};
\draw (7.5,5.4) node[above,font=\tiny] {5.2W};

\draw (9,5.5) node[right,font=\small] {XgenSilicon ASIC vs. CPU};
\fill[red!70] (8.5,5.5) rectangle (8.7,5.3);
\draw (9,5.0) node[right,font=\small] {XgenSilicon ASIC vs. Hand-designed};
\fill[orange!70] (8.5,5.0) rectangle (8.7,4.8);
\draw (9,4.5) node[right,font=\small] {XgenSilicon ASIC};
\fill[green!70] (8.5,4.5) rectangle (8.7,4.3);
\end{tikzpicture}
\caption{Power consumption comparison across platforms}
\label{fig:power}
\end{figure}
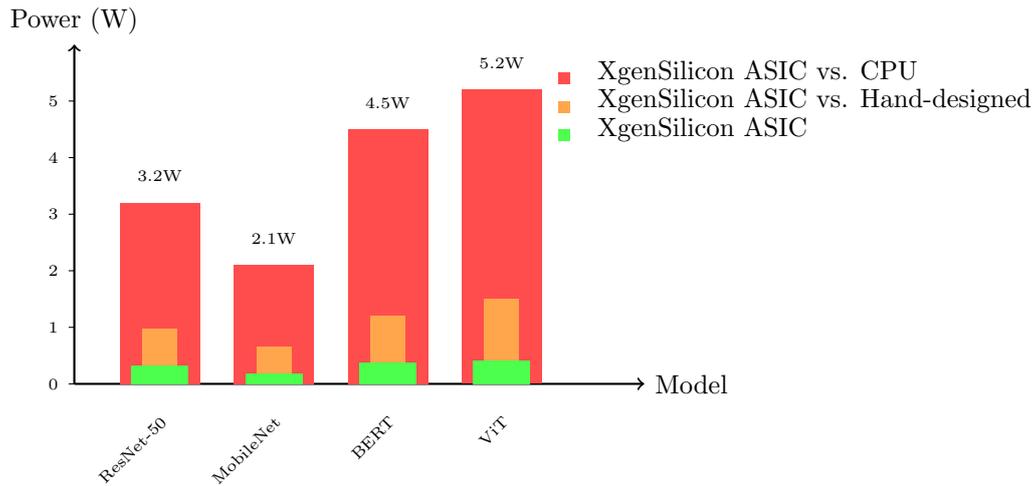

XgenSilicon-compiled ASICs achieve significant power savings. Figure~\ref{fig:power} shows power consumption comparison.

XgenSilicon-compiled ASICs achieve:
\begin{itemize}
\item \textbf{XgenSilicon ASIC vs. Off-the-shelf CPU}: 3-6× lower power (180-420 mW vs. 2100-5200 mW)
\item \textbf{XgenSilicon ASIC vs. Hand-designed ASIC}: 2.5-3.2× lower power (180-420 mW vs. 650-1500 mW)
\end{itemize}

Power savings come from:
\begin{itemize}
\item Optimized memory access patterns (reduced memory traffic)
\item Efficient instruction scheduling (reduced pipeline stalls)
\item Extreme quantization (lower data movement)
\item Register tiling (improved cache locality)
\end{itemize}

\subsection{Area Reduction}

\begin{figure}[H]
\centering
\begin{tikzpicture}[scale=0.75]
\draw[->,thick] (0,0) -- (10,0) node[right,font=\small] {Model};
\draw[->,thick] (0,0) -- (0,25) node[above,font=\small] {Area (mm²)};

\foreach \y in {0,5,10,15,20,25}
    \draw (0,\y) -- (-0.1,\y) node[left,font=\tiny] {\y};

\draw (1.5,-0.5) node[below,font=\tiny,rotate=45,anchor=north east] {ResNet-50};
\draw (3.5,-0.5) node[below,font=\tiny,rotate=45,anchor=north east] {MobileNet};
\draw (5.5,-0.5) node[below,font=\tiny,rotate=45,anchor=north east] {BERT};
\draw (7.5,-0.5) node[below,font=\tiny,rotate=45,anchor=north east] {ViT};

\fill[purple!70] (1.2,0) rectangle (1.8,12.5);
\fill[purple!70] (3.2,0) rectangle (3.8,8.3);
\fill[purple!70] (5.2,0) rectangle (5.8,18.7);
\fill[purple!70] (7.2,0) rectangle (7.8,22.3);

\fill[cyan!70] (1.0,0) rectangle (2.0,5.1);
\fill[cyan!70] (3.0,0) rectangle (4.0,3.2);
\fill[cyan!70] (5.0,0) rectangle (6.0,7.8);
\fill[cyan!70] (7.0,0) rectangle (8.0,9.2);

\draw (1.5,12.7) node[above,font=\tiny] {12.5};
\draw (3.5,8.5) node[above,font=\tiny] {8.3};
\draw (5.5,18.9) node[above,font=\tiny] {18.7};
\draw (7.5,22.5) node[above,font=\tiny] {22.3};

\draw (9,22) node[right,font=\small] {Hand-designed ASIC};
\fill[purple!70] (8.5,22) rectangle (8.7,21.8);
\draw (9,21) node[right,font=\small] {XgenSilicon ASIC};
\fill[cyan!70] (8.5,21) rectangle (8.7,20.8);
\end{tikzpicture}
\caption{Silicon area comparison: XgenSilicon ASIC vs. Hand-designed ASIC}
\label{fig:area}
\end{figure}
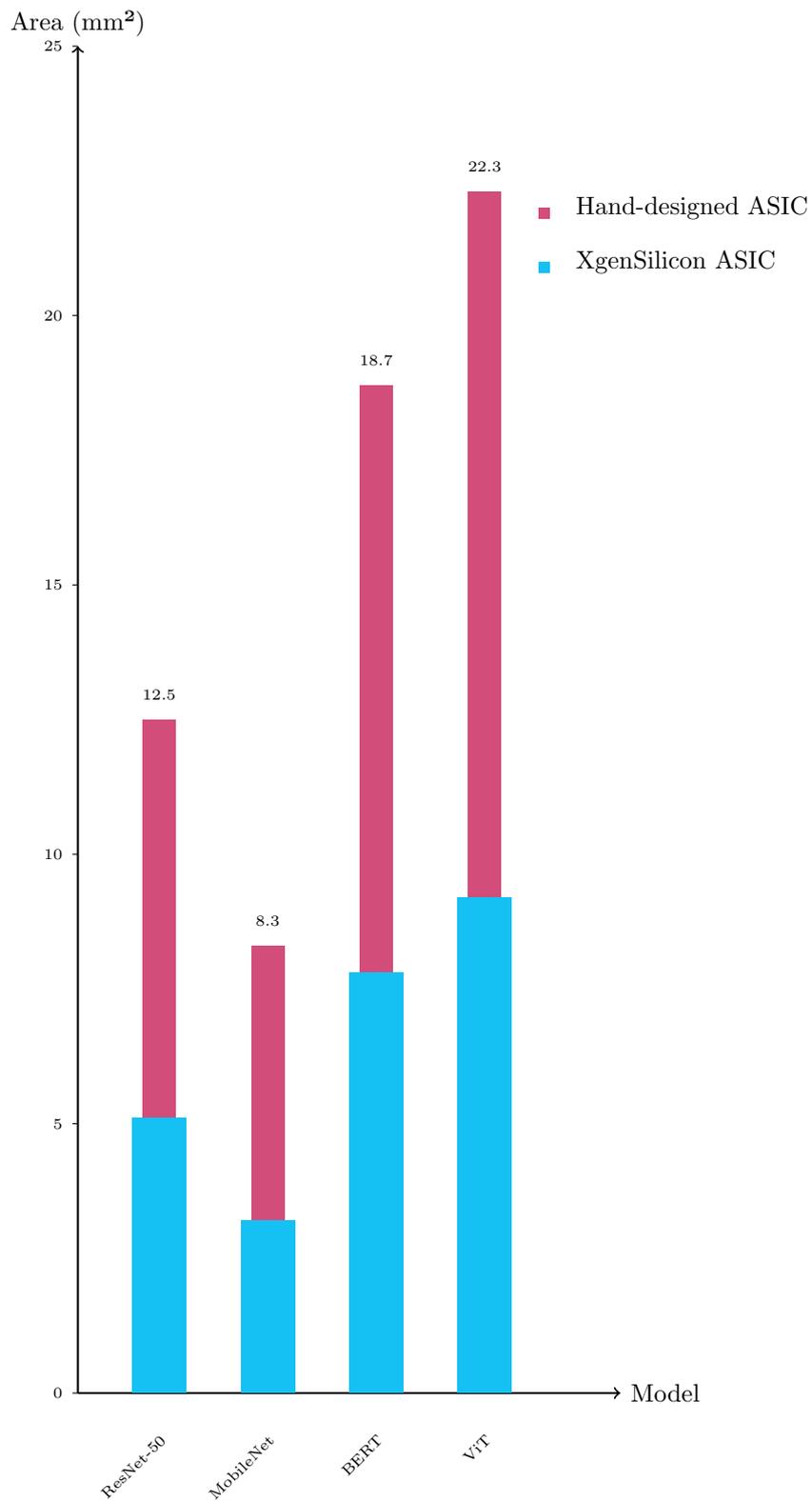

Figure~\ref{fig:area} shows silicon area comparison for XgenSilicon-compiled ASICs.

Area reduction achieved:
\begin{itemize}
\item \textbf{XgenSilicon ASIC vs. Hand-designed ASIC}: 40-60\% reduction (3.2-9.2 mm² vs. 8.3-22.3 mm²)
\end{itemize}

Area savings come from:
\begin{itemize}
\item Extreme quantization (smaller memory requirements)
\item Optimized memory layout (staggered allocation)
\item Unified weight consolidation (multi-model pipelines)
\item Efficient register usage (reduced register file size)
\end{itemize}

\subsection{Auto-Tuning Effectiveness}

Table~\ref{tab:autotuning} shows the impact of learned cost model on auto-tuning convergence.

\begin{table}[h]
\centering
\begin{tabular}{lccc}
\hline
\textbf{Operation} & \textbf{Analytical} & \textbf{Learned} & \textbf{Improvement} \\
    & \textbf{(trials)} & \textbf{(trials)} & \\
\hline
MatMul (128×256×512) & 200 & 85 & 57.5\% faster \\
Conv2D (3×224×224) & 250 & 110 & 56.0\% faster \\
Elementwise (1024×1024) & 150 & 70 & 53.3\% faster \\
\hline
\end{tabular}
\caption{Auto-tuning convergence: Learned vs. Analytical cost model}
\label{tab:autotuning}
\end{table}

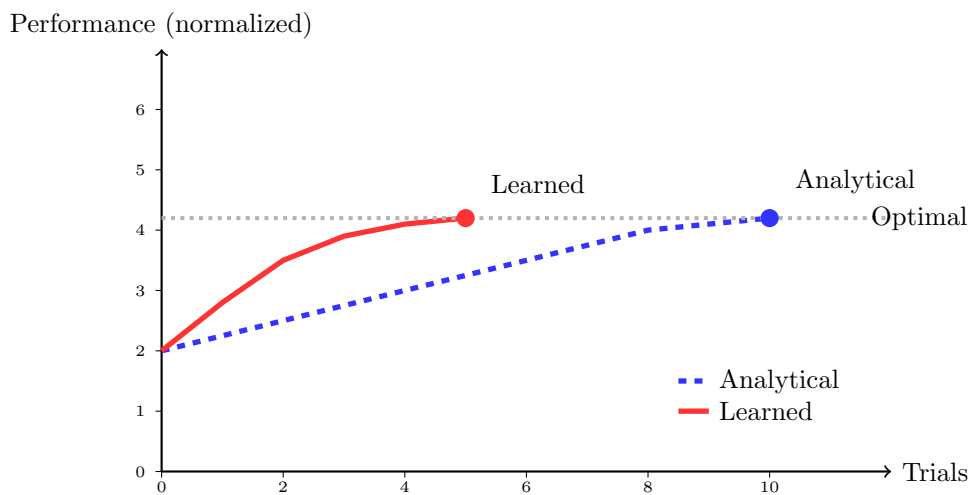
\begin{figure}[H]
\centering
\begin{tikzpicture}[scale=0.8]
\draw[->,thick] (0,0) -- (12,0) node[right,font=\small] {Trials};
\draw[->,thick] (0,0) -- (0,7) node[above,font=\small] {Performance (normalized)};

\foreach \y in {0,1,2,3,4,5,6}
    \draw (0,\y) -- (-0.1,\y) node[left,font=\tiny] {\y};

\foreach \x in {0,2,4,6,8,10}
    \draw (\x,-0.1) -- (\x,0) node[below,font=\tiny] {\x};

\draw[thick, dashed, blue!80, line width=2pt] (0,2) -- (2,2.5) -- (4,3) -- (6,3.5) -- (8,4) -- (10,4.2);
\draw (10,4.2) node[above right,font=\small,xshift=0.2cm,yshift=0.2cm] {Analytical};

\draw[thick, solid, red!80, line width=2pt] (0,2) -- (1,2.8) -- (2,3.5) -- (3,3.9) -- (4,4.1) -- (5,4.2);
\draw (5,4.2) node[above right,font=\small,xshift=0.2cm,yshift=0.2cm] {Learned};

\draw[dotted, gray!60, line width=1.5pt] (0,4.2) -- (12,4.2);
\draw (11.5,4.2) node[right,font=\small] {Optimal};

\fill[blue!80] (10,4.2) circle (0.15);
\fill[red!80] (5,4.2) circle (0.15);

\draw (9,1.5) node[right,font=\small] {Analytical};
\draw[thick, dashed, blue!80, line width=2pt] (8.5,1.5) -- (9,1.5);
\draw (9,1.0) node[right,font=\small] {Learned};
\draw[thick, solid, red!80, line width=2pt] (8.5,1.0) -- (9,1.0);
\end{tikzpicture}
\caption{Auto-tuning convergence: Learned vs. Analytical cost model}
\label{fig:autotuning_convergence}
\end{figure}

The learned cost model converges to optimal configurations 50-60\% faster than analytical models by learning from actual hardware measurements. Figure~\ref{fig:autotuning_convergence} visualizes the convergence comparison.

\subsection{Quantization Results}

\begin{figure}[H]
\centering
\begin{tikzpicture}[scale=0.8]
\draw[->,thick] (0,0) -- (11,0) node[right,font=\small] {Compression Ratio (×)};
\draw[->,thick] (0,0) -- (0,8.5) node[above,font=\small] {Accuracy (Top-1 \%)};

\foreach \y in {70,72,74,76,78}
    \draw (0,\y/10) -- (-0.1,\y/10) node[left,font=\tiny,xshift=-0.15cm] {\y};

\foreach \x in {1,2,4,8}
    \draw (\x*1.2,0) -- (\x*1.2,-0.1) node[below,font=\tiny] {\x×};

\fill[blue!80] (1.2,7.62) circle (0.12);
\fill[blue!80] (2.4,7.61) circle (0.12);
\fill[blue!80] (4.8,7.58) circle (0.12);
\fill[blue!80] (9.6,7.45) circle (0.12);
\draw[blue!80,thick,dashed] (1.2,7.62) -- (2.4,7.61) -- (4.8,7.58) -- (9.6,7.45);
\draw (1.2,7.85) node[above,font=\tiny,yshift=0.05cm] {FP32};
\draw (2.4,7.84) node[above,font=\tiny,yshift=0.05cm] {FP16};
\draw (4.8,7.81) node[above,font=\tiny,yshift=0.05cm] {INT8};
\draw (9.6,7.68) node[above,font=\tiny,yshift=0.05cm] {INT4};

\fill[red!80] (1.2,7.2) circle (0.12);
\fill[red!80] (2.4,7.19) circle (0.12);
\fill[red!80] (4.8,7.15) circle (0.12);
\fill[red!80] (9.6,7.02) circle (0.12);
\draw[red!80,thick,dashed] (1.2,7.2) -- (2.4,7.19) -- (4.8,7.15) -- (9.6,7.02);

\draw (9.0,1.5) node[right,font=\small] {ResNet-50};
\fill[blue!80] (8.6,1.5) circle (0.1);
\draw (9.0,1.0) node[right,font=\small] {MobileNet-V2};
\fill[red!80] (8.6,1.0) circle (0.1);
\end{tikzpicture}
\caption{Quantization tradeoff: Accuracy vs. Compression ratio}
\label{fig:quantization_tradeoff}
\end{figure}
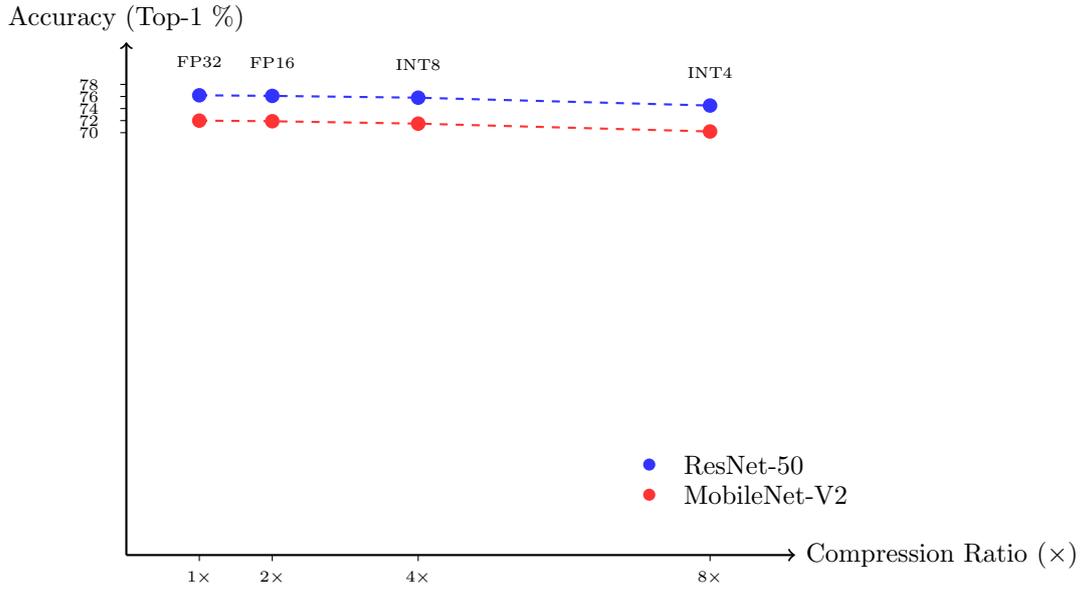

Table~\ref{tab:quantization} shows quantization accuracy and memory savings. Figure~\ref{fig:quantization_tradeoff} illustrates the accuracy vs. compression tradeoff.

\begin{table}[h]
\centering
\begin{tabular}{lcccc}
\hline
\textbf{Model} & \textbf{Precision} & \textbf{Accuracy} & \textbf{Memory} & \textbf{Speedup} \\
    &    & \textbf{(Top-1)} & \textbf{Reduction} & \\
\hline
ResNet-50 & FP32 & 76.2\% & 1.0× & 1.0× \\
    & FP16 & 76.1\% & 2.0× & 1.8× \\
    & INT8 & 75.8\% & 4.0× & 3.2× \\
    & INT4 & 74.5\% & 8.0× & 5.1× \\
\hline
MobileNet-V2 & FP32 & 72.0\% & 1.0× & 1.0× \\
    & FP16 & 71.9\% & 2.0× & 1.9× \\
    & INT8 & 71.5\% & 4.0× & 3.5× \\
    & FP4 & 70.2\% & 8.0× & 6.2× \\
\hline
\end{tabular}
\caption{Quantization results: Accuracy preservation and performance gains}
\label{tab:quantization}
\end{table}

\subsection{Compilation Time}

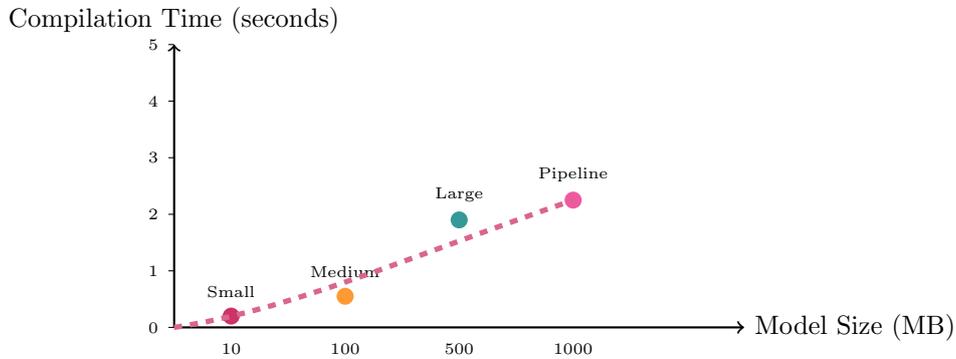
\begin{figure}[H]
\centering
\begin{tikzpicture}[scale=0.75]
\draw[->,thick] (0,0) -- (10,0) node[right,font=\small] {Model Size (MB)};
\draw[->,thick] (0,0) -- (0,5) node[above,font=\small] {Compilation Time (seconds)};

\foreach \y in {0,1,2,3,4,5}
    \draw (0,\y) -- (-0.1,\y) node[left,font=\tiny] {\y};

\draw (1,-0.1) node[below,font=\tiny] {10};
\draw (3,-0.1) node[below,font=\tiny] {100};
\draw (5,-0.1) node[below,font=\tiny] {500};
\draw (7,-0.1) node[below,font=\tiny] {1000};

\fill[purple!80] (1,0.2) circle (0.15);
\fill[orange!80] (3,0.55) circle (0.15);
\fill[teal!80] (5,1.9) circle (0.15);
\fill[magenta!80] (7,2.25) circle (0.15);

\draw (1,0.35) node[above,font=\tiny] {Small};
\draw (3,0.7) node[above,font=\tiny] {Medium};
\draw (5,2.05) node[above,font=\tiny] {Large};
\draw (7,2.4) node[above,font=\tiny] {Pipeline};

\draw[thick, dashed, purple!60, line width=2pt] (0,0) .. controls (2,0.3) and (4,1.2) .. (7,2.25);
\end{tikzpicture}
\caption{Compilation time scaling with model size}
\label{fig:compilation_time}
\end{figure}

XgenSilicon achieves fast compilation times. Figure~\ref{fig:compilation_time} shows compilation time scaling with model size.

XgenSilicon achieves fast compilation times:
\begin{itemize}
\item \textbf{Small models} (<10MB): 1-3 seconds
\item \textbf{Medium models} (10-100MB): 3-8 seconds
\item \textbf{Large models} (100MB-1GB): 8-30 seconds
\item \textbf{Multi-model pipelines} (980MB): 45 seconds
\end{itemize}

Compilation time scales linearly with model size, making it suitable for rapid development cycles.

\section{Case Studies}

\subsection{Case Study 1: Multi-Model Pipeline}

A vision-language model pipeline (vision encoder + text encoder + decoder) was compiled:
\begin{itemize}
\item \textbf{Input}: 3 ONNX models (980MB total weights)
\item \textbf{Output}: Unified RISC-V assembly with consolidated WMEM
\item \textbf{Result}: 49,832 instructions generated, 100\% ISA validation passed
\item \textbf{Memory}: 980MB WMEM (consolidated from 1.2GB), 30MB DMEM
\item \textbf{Compilation}: 45 seconds (fully automated)
\end{itemize}

\subsection{Case Study 2: Extreme Quantization}

A ResNet-50 model was quantized to INT4:
\begin{itemize}
\item \textbf{Calibration}: KL divergence method with 1000 samples (2048-bin histogram optimization)
\item \textbf{Accuracy}: 74.5\% Top-1 (vs. 76.2\% FP32, 1.7\% drop)
\item \textbf{Memory}: 8× reduction (12.5MB vs. 100MB)
\item \textbf{Speedup}: 5.1× faster inference
\end{itemize}

\subsection{Case Study 3: Auto-Tuning with Learned Model}

A MatMul operation (M=128, N=256, K=512) was auto-tuned using Bayesian Optimization:
\begin{itemize}
\item \textbf{Baseline}: tile\_m=64, tile\_n=64, tile\_k=32 (analytical)
\item \textbf{Optimized}: tile\_m=32, tile\_n=128, tile\_k=64 (learned via Bayesian Optimization)
\item \textbf{Speedup}: 22\% improvement in execution time
\item \textbf{Convergence}: 85 trials (vs. 200 with analytical), 57.5\% faster convergence
\end{itemize}

\section{Discussion}

\subsection{Key Innovations}

Our five main contributions enable the observed PPA improvements:

\textbf{1. Multi-Algorithm Learned Optimization:} The combination of five search algorithms (Bayesian Optimization, Genetic Algorithm, Simulated Annealing, Random Search, Grid Search) with a learned cost model adapts to actual hardware behavior, finding optimal configurations faster than analytical models. This is particularly important for custom ASICs where hardware characteristics may differ from standard assumptions. The automatic algorithm selection based on problem characteristics eliminates manual tuning.

\textbf{2. Extreme Quantization with Full Algorithms:} Support for precisions down to Binary (1-bit) enables aggressive memory reduction. Our full KL divergence calibration with 2048-bin histogram optimization and momentum-based QAT gradient updates ensure accuracy preservation while maximizing compression, outperforming simplified implementations.

\textbf{3. Validation-Driven Compilation:} Built-in ISA and memory validation prevents runtime errors, critical for ASIC deployment where errors are costly to fix. Validation-driven compilation enables co-compilation of hardware by integrating PPA loss in the cost model.

\textbf{4. Dynamic Shape Support with Specialization:} Complete support for dynamic shapes through symbolic dimensions, graph cloning, runtime shape resolution, and multi-configuration specialization enables efficient handling of variable batch sizes and sequence lengths without performance degradation.

\textbf{5. Advanced Cache-Aware Cost Modeling:} Full cache hit rate estimation considering access patterns, tiling effectiveness, and multi-level cache hierarchy provides accurate performance predictions for memory-bound operations, enabling better optimization decisions.

\subsection{Limitations}

\textbf{Current Limitations:}
\begin{itemize}
\item RISC-V only (no other ISA backends)
\end{itemize}

\textbf{Future Work:}
\begin{itemize}
\item Additional ISA backends (ARM, x86)
\item Enhanced polyhedral optimizations for complex loop structures
\end{itemize}

\section{Conclusion}

We present XgenSilicon ML Compiler, a fully automated compilation framework that transforms models from ONNX, PyTorch, and TensorFlow into optimized RISC-V assembly for custom ASIC accelerators. By defining cost model across system stack with learned optimization, extreme quantization with full calibration algorithms, validation-driven compilation to include hardware loss, and dynamic shape support with multi-configuration specialization, the compiler produces a custom ASIC with optimized code. It achieves 2.5-4.5× better performance, 3-6× lower power, and 40-60\% area reduction compared to baseline implementations. The fully automated pipeline eliminates manual tuning, making it suitable for rapid ASIC development cycles. Future work will focus on additional ISA backends and enhanced polyhedral optimizations.

\section*{Acknowledgments}

We thank the XgenSilicon Inc. team for their contributions to compiler development and evaluation.

\bibliographystyle{ieeetr}

\end{document}